\documentclass[a4paper,fleqn]{cas-dc}

\usepackage[numbers]{natbib}

\def\tsc#1{\csdef{#1}{\textsc{\lowercase{#1}}\xspace}}
\tsc{WGM}

\begin{document}
\let\WriteBookmarks\relax
\def\floatpagepagefraction{1}
\def\textpagefraction{.001}

\shorttitle{Resonance with self-organized energy transfer}

\shortauthors{Optomechanical resonance Lin et~al.}

\title [mode = title]{Nonlinear optomechanical resonance entering a self-organized energy transfer pattern}                      



%
\author[1]{Qing Lin} [orcid=0000-0002-1346-0167]

\cormark[1]





\affiliation[1]{organization={Fujian Key Laboratory of Light Propagation and Transformation \& Institute of Systems Science, 
College of Information Science and Engineering, Huaqiao University},
    addressline={}, 
    city={Xiamen},
    postcode={361201}, 
    country={China}}

\author[1]{Yi Wu}[]
\cormark[1]

\author[1]{Gang Li}[]




\author%
[2]
{Bing He}[orcid=0000-0002-7169-0446]
\cormark[2]
\ead{bing.he@umayor.cl}

\affiliation[2]{organization={Center for Quantum Optics and Quantum Information, Universidad Mayor},
    addressline={Camino La Pir\'{a}mide 
5750}, 
    city={Huechuraba},
    state={RM},
    country={Chile}}

\cortext[cor1]{These authors contribute equally to this work}
\cortext[cor2]{corresponding author}



\begin{abstract}
The energy transfer between different subsystems or different vibration modes is always one of the most interested problems in the study of the resonance phenomena in coupled nonlinear dynamical systems. With an optomechanical system operating in the regime of unresolved sideband, where its mechanical frequency is lower than the cavity field damping rate, we illustrate the existence of a special nonlinear resonance phenomenon. This type of previously unknown resonance manifests an organized pattern of the coupled cavity field and mechanical oscillation, so that the cavity field precisely pushes the mechanical oscillator within an appropriate small time window in each mechanical oscillation period and the mechanical energy will increase by a jump of almost fixed amount after each oscillation cycle. The scenario is realized at a resonance point where the frequency difference of two driving fields matches the mechanical frequency of the system, and this condition of drive-frequency match is found to trigger a mechanism to lock the two subsystems of an unresolved-sideband optomechanical system into a highly ordered energy transfer as the above mentioned. Due to a significantly enhanced nonlinearity in the vicinity of the resonance point, optical frequency combs can be generated under pump powers of thousand times lower, as compared to the use of a single-tone driving field for the purpose. An unresolved sideband system under the drives without satisfying the resonance condition also demonstrates other interesting dynamical behaviors. Most of all, by providing a realistic picture for the nonlinear optomechanical dynamics in unresolved sideband regime, our study points to a direction to observe novel dynamical phenomena and realize other applications with the systems of less technical restrictions.    
\end{abstract}




\maketitle

\section{Introduction}
Resonance phenomena are ubiquitous in both nature and artificial structures, and refer to the maximum response of a vibration to external oscillatory forces. Beyond the scenarios of a single oscillator, such as the well-known stochastic resonance \cite{gammaitoni1998stochastic} and parametric resonance \cite{fossen2011parametric}, the nonlinear resonance phenomena in coupled systems were less studied but are more important to applications. For example, {\it internal resonance} in coupled structures \cite{jackson1963nonlinear,ford1963computer,manevitch2005mechanics,bajaj2018internal} is highly interested and useful for its energy transfer between different vibration modes under the condition that their frequencies are in some proper ratios. The current work is about a nonlinear resonance in optomechanical systems (OMS), which manifests a unique pattern of energy transfer between coupled cavity field and mechanical oscillator. 

In addition to their potential values in quantum technology \cite{aspelmeyer2014cavity,barzanjeh2022optomechanics}, OMSs provide a good platform to study nonlinear dynamics. Nonlinear dynamical behaviors in OMS were previously known as the bifurcations into limit cycles \cite{rokhsari2005radiation,marquardt2006dynamical,zaitsev2011forced,krause2015nonlinear,colombano2019synchronization,sheng2020self,lin2021catastrophic,wang2021passive} and chaos \cite{carmon2005temporal,carmon2007chaotic,ma2014formation,bakemeier2015route,zhu2022cavity}. The dynamical scenarios in the first category can be summarized by the response of an OMS to the external driving fields of varied frequencies and intensities. In the regime where the systems' mechanical frequencies $\omega_m$ exceed their optical cavity damping rates $\kappa$, a driving laser will turn an OMS from a steady state to an oscillation when its power is above a threshold of Hopf bifurcation. Direct numerical calculations based on the full nonlinear dynamics \cite{he2023dynamical} show that, along the way of scanning the frequency of a driving laser of the same power, the induced mechanical oscillation has the largest amplitude at where the drive frequency is blue-detuned from the system's resonant cavity frequency exactly by the mechanical frequency $\omega_m$. This is a resonance point of the same dynamical pattern of harmonic oscillation, commonly seen in nonlinear dynamical systems. Parallel to this category of nonlinear dynamics, most recent researches on the regime $\omega_m/\kappa>1$ concern the side of quantum mechanics, such as mechanical ground states 
(see, e.g. \cite{schliesser2009resolved,teufel2011sideband,chan2011laser}) and others (see, e.g. \cite{purdy2013strong,aggarwal2020room}), which can be well described by the linearized dynamics around a steady state \cite{aspelmeyer2014cavity} or by another quantum dynamical approach \cite{he2023dynamical,he2017radiation}. To demonstrate quantum properties, an OMS should be ideally fabricated to meet the resolved sideband condition $\omega_m\gg \kappa$, and this requirement poses a technical challenge of tremendously increasing the optical finesse and limits the mass/size of the mechanical element of an OMS. Instead, the phenomena described in the current work must exist in another regime where $\omega_m<\kappa$. Most previous experiments in such unresolved sideband regime \cite{doolin2014nonlinear,brawley2016nonlinear,leijssen2017nonlinear,meng2022measurement} were performed in the bad cavity limit, to assume a relatively trivial picture of steady intra-cavity photon number $|a|^2$ evolved under 
a large damping rate $\kappa$, though there have been many interesting theoretical studies involving the regime, such as the relaxation of the requirement $\omega_m\gg\kappa$ for realizing quantum effects (see. e.g. \cite{ojanen2014ground,bennett2016quantum,neumeier2018exploring,zhang2019strong,han2019mechanical,shahandeh2019optomechanical,zhang2020large,kanari2022two}) and the improvement of quantum efficiencies with multi-mode OMSs (see, e.g. \cite{lai2018simultaneous, lai2021domino, xu2022millionfold, lai2022efficient}). A nontrivial scenario of nonlinearity known in unresolved sideband regime is the creation of pulsed cavity field under stronger driving fields \cite{poot2012backaction,miri2018optomechanical,xu2021chip}. However, as we will learn from the later discussion, no resonance phenomenon exists for any unresolved-sideband OMS driven by an optical field of single frequency.

Our concerned resonance phenomenon emerges when a suitable unresolved-sideband OMS is driven by two fields (or a two-tone field) with the difference of their frequencies matching the mechanical frequency of the system. At this resonance point, the system exhibits a totally different dynamical pattern of boundless energy harvesting from any other dynamical pattern at a point off the resonance. Boundless energy harvesting in spite of a nonzero damping rate for an oscillator was known to be possible by means of parametric resonance \cite{fossen2011parametric} or both classical \cite{veksler1945new,mcmillan1945synchrotron} and quantum version \cite{marcus2004quantum,barth2014quantum} of {\it autoresonance} (see \cite{rajasekar2016nonlinear} for an overview) under a chirped driving force (its frequency is a function of time) which brings the oscillator to higher energy constantly. We realize a boundless energy harvesting through a different mechanism. 
It is found that the difference of two drive tones works as a control parameter that can trigger a mechanism of organizing the motions of two subsystems (the cavity field and mechanical oscillator) of an OMS. Once this parameter is tuned to be close to the mechanical frequency of an unresolved-sideband OMS, the system will enter a pattern of cooperative energy transfer between its two subsystems, so that this classical dynamical system exhibits a repeatedly "quantized" energy addition of almost fixed amount in each mechanical oscillation cycle. The realized step-like energy increment to the mechanical oscillator is even more regular than those in 
a corresponding process of boundless energy harvesting 
through a quantum {\it autoresonance} \cite{marcus2004quantum,barth2014quantum}. This previously unknown mechanism has a capability of significantly lowering the power to generate the optical pulse trains applied in communications and precise measurements.

The rest of the paper is organized as follows. After an illustration of the dynamical model based on the realistic systems in Sec. 2, 
we present in Sec. 3 the detailed discussion on the features of the concerned phenomenon and the existence of a mechanism behind the phenomenon. In the following section, Sec. 4, we describe the phenomena when the system parameters are modified to lose the mentioned resonance. Some other issues, such as an interesting feature of preserving the resonance under the parallel shifts of two drive tones, the perturbation from thermal noise, and the effects of the phase mismatch between two drives, are discussed in Sec. 5. By Sec. 6 and Sec. 7, respectively, we provide a feasible application of the resonance phenomenon and provide necessary information about the experimental implementation of this dynamical scenario, before we conclude the work with the last section.

\begin{figure}
	\centering
		\includegraphics[scale=.34]{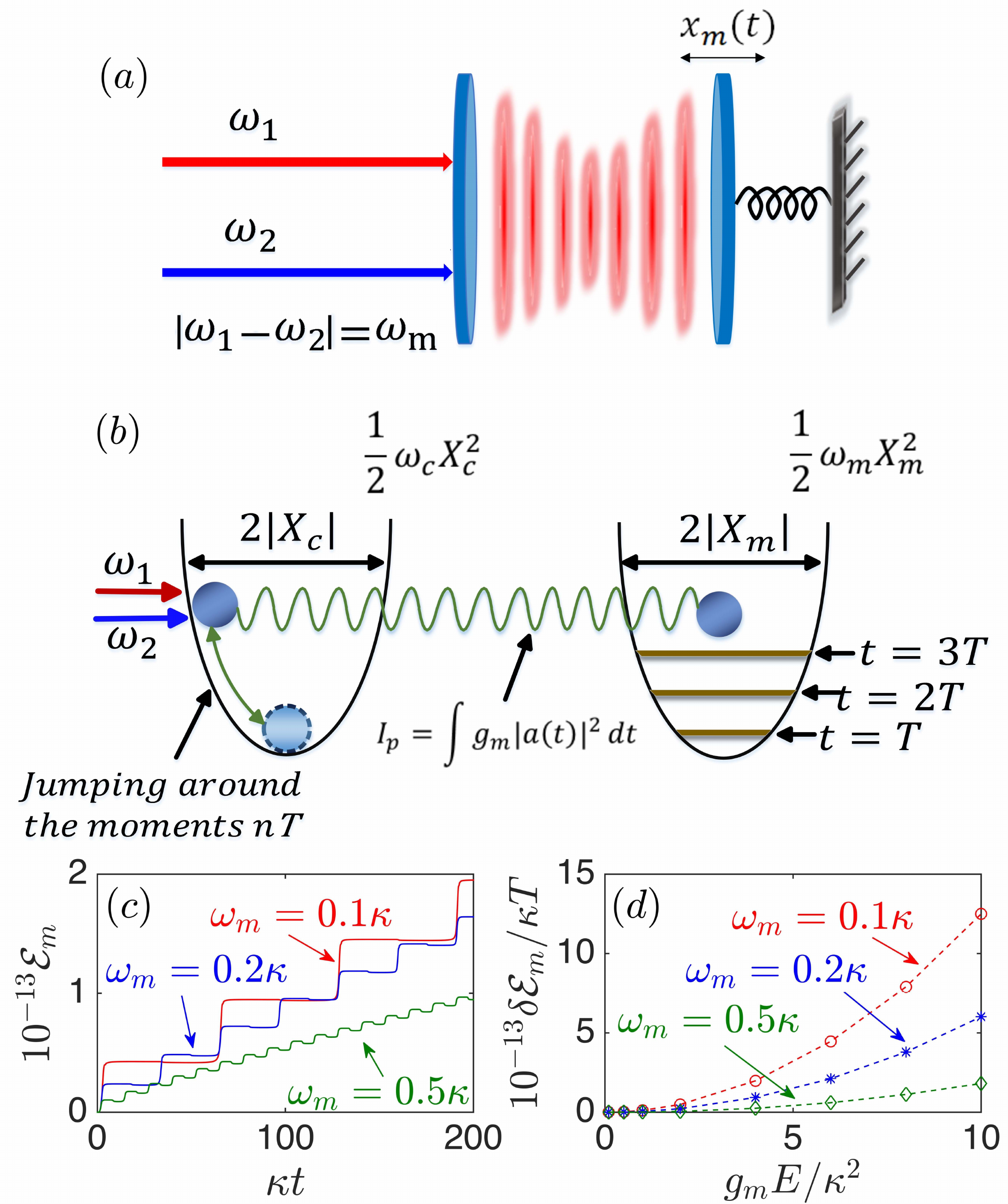}
	\caption{{\bf Model and phenomenon.} (a) An exemplary setup driven by the fields under a frequency condition. 
(b) The model of two oscillators respectively trapped in the potential well $\frac{1}{2}\hbar\omega_{c(m)}X^2_{c(m)}$ and coupled through the potential in Eq. (\ref{potential}). 
When two driving fields are tuned to a resonance condition, the repeated excitation of the first one around each moment $nT$ becomes highly coordinated with the motion of the second, which is like being constantly excited to the equally distanced energy levels after each period $T$, though the system dynamics is classical by nature. (c) The nonlinear resonance. All simulated processes are from the initial condition $X_m (P_m)=0$, $|a|^2=0$, and are obtained with $g_m/\kappa=10^{-5}$, $\omega_m/\gamma_m=10^6$, $E/\kappa=2\times 10^5$, $\omega_1=\omega_c$, and $\omega_2=\omega_c-\omega_m$. No specific $\omega_c$ is necessary by working with 
Eq. (\ref{eq}). (d) The stair slope tendency with the ratio $g_mE/\kappa^2$ ($g_m$ is fixed), for the mechanical frequencies used in (c).}
	\label{FIG:1}
 \vspace{-0.3cm}
\end{figure}

\section{Model}
We illustrate with a Fabry-Perot-type OMS in Fig. \ref{FIG:1}(a). Under the radiation pressure from the cavity photons with the number $|a|^2$, the energy $\hbar\omega_c|a|^2$ stored between two mirrors, which form a cavity with its resonance frequency $\omega_c$, will be modified to $\hbar\omega_c(1-x_m/L)|a|^2$ after a tiny displacement $x_m(t)=\sqrt{\hbar/(m\omega_m)}X_m(t)$ ($X_m(t)$ is the dimensionless displacement) of the end mirror with an effective mass $m$. This modification is valid under the condition $|x_m|\ll L$ (the size of the cavity), and it gives rise 
to an interaction potential 
\begin{eqnarray}
V_{int}(t)=-\hbar g_mX_m(t)|a(t)|^2
\label{potential}
\end{eqnarray}
between the cavity field and mechanical mode, where $g_m=\sqrt{\frac{\hbar}{m\omega_m}}/L\times \omega_c$ is the optomechanical coupling constant.
Then, by modeling the cavity field as another oscillator with its two perpendicular quadratures being the dimensionless displacement and momentum ($X_c$ and $P_c$), such that $a=(X_c+iP_c)/\sqrt{2}$, one has the nonlinear dynamical equations 
\begin{eqnarray}
&&\dot{a}=-\kappa a-i(\omega_c-g_mX_m)a+E(e^{-i\omega_1 t+i\theta_1}+e^{-i\omega_2 t+i\theta_2})\nonumber\\
&&\dot{X}_m=\omega_mP_m,\nonumber\\
&&\dot{P}_m=-\gamma_mP_m-\omega_mX_m+g_m|a|^2+\sqrt{2\gamma_m}\xi_m(t)
\label{2}
\end{eqnarray}
for the model in Fig. \ref{FIG:1}(b), where $\gamma_m$ is the mechanical damping rate, $\omega_{1(2)}$ the drive frequency, and the used dimensionless mechanical momentum $P_m(t)$ is related to the real momentum $p_m(t)$ as $p_m(t)=\sqrt{\hbar m\omega_m}P_m(t)$. The cavity field damping rate 
$\kappa=\kappa_e+\kappa_i$ includes two parts; $\kappa_e$ measures the coupling to the driving field and $\kappa_i$ indicates the intrinsic loss. In the regime where $\omega_m>\kappa$, the pump power $\hbar\omega_{1(2)}E^2/(2\kappa_e)$ should be higher than a threshold of Hopf bifurcation to drive an OMS to oscillate. Before Sec. 5 we neglect the thermal noise term $\sqrt{2\gamma_m}\xi_m(t)$ and assume an identical driving field phase $\theta_1=\theta_2=0$.

By a transformation $a\rightarrow ae^{-i\omega_c t}$ of the cavity field mode, the first equation in Eq. (\ref{2}) will take the form
\begin{eqnarray}
&&\dot{a}=-\kappa a+ig_mX_ma+E\left(e^{i\Delta_1 t}+e^{i\Delta_2 t}\right)
\label{eq}
\end{eqnarray}
in the reference frame rotating at the cavity frequency $\omega_c$, where $\Delta_{1(2)}=\omega_c-\omega_{1(2)}$ is the detuning of one drive tone from the resonant frequency $\omega_c$. It will not modify the other equations in Eq. (\ref{2}). Then, no specific resonant cavity $\omega_c$ will be needed in the numerical integral of the dynamical equations, while the damping rate $\kappa$ is used to scale all other parameters in such equations reformatted with a dimensionless evolution time $\kappa t$. 

In the numerical calculations we adopt the mechanical quality factor $Q=\omega_m/\gamma_m$ in the range $10^{3}-10^{6}$, the optomechanical coupling constant $g_m=10^{-5}\kappa$, and the mechanical frequency in the order of $\omega_m\sim 0.1\kappa$. If the optical damping rate $\kappa$ is in the order of $2\pi\times 10$ MHz, the required coupling $g_m$ is only in the order of $2\pi \times 100$ Hz. These parameters are well within those achieved in the past experiments reviewed in Ref. \cite{aspelmeyer2014cavity}. Because our concerned systems are of unresolved sideband, the requirements on their fabrication and performance are much less demanding. More details in this aspect are given in Sec. 6 and Sec. 7. 

\begin{figure}
	\centering
		\includegraphics[scale=.05]{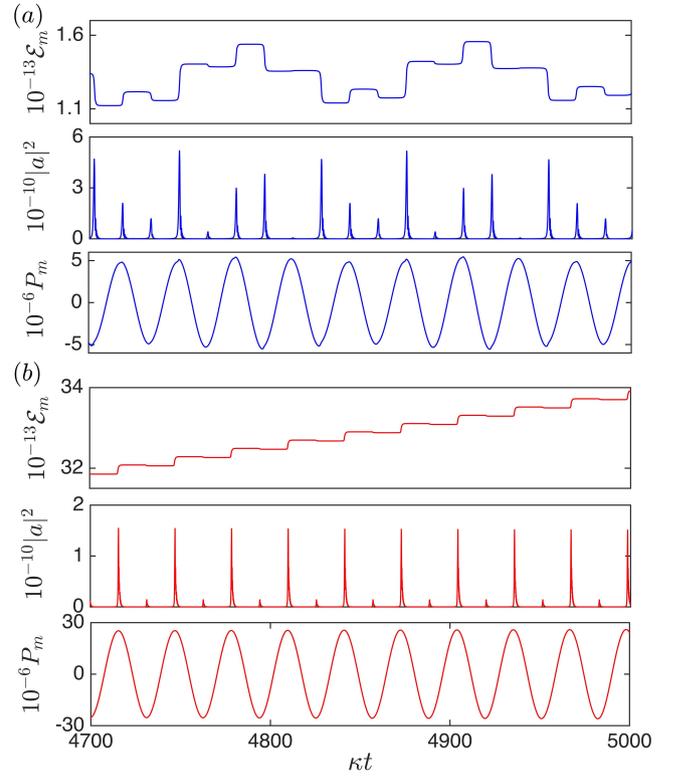}
	\caption{{\bf Comparison between two scenarios.} 
(a) The simultaneously evolving mechanical energy, cavity photon number, 
and mechanical momentum. Under the continuous pumping there is only an averaged $|a|^2\sim 10^4$ 
(invisible by the used scale) between two pulse peaks. The used OMS has $\omega_m/\kappa=0.2$, $\gamma_m/\omega_m=10^{-6}$, and $g_m/\kappa=10^{-5}$, while the pumps are with $E/\kappa=2\times 10^5$ 
and $\omega_1=\omega_c$, $\omega_2=\omega_c-0.25\kappa$. 
(b) The corresponding evolution processes due to a difference $\omega_2=\omega_c-0.2\kappa$. 
Here, the mechanical momentum $P_m$ evolves smoothly and an obvious change of its amplitude should be seen over a longer evolution duration $\kappa\Delta t$.}
	\label{FIG:2}
\end{figure}

\begin{figure}
	\centering
		\includegraphics[scale=.05]{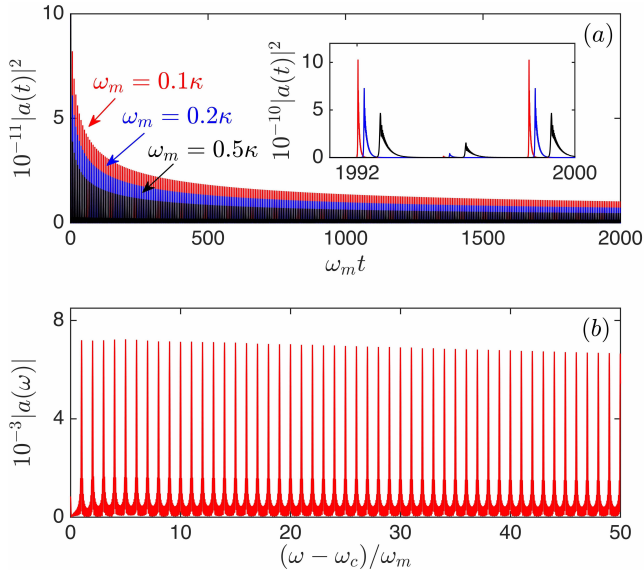}
	\caption{{\bf Generated pulses from drive-frequency match.} (a) The pulses of the exemplary systems 
[$\omega_m/\kappa=0.1$ (red), $0.2$ (blue) and $0.5$ (black)], which drop at the same pace after each period $T$. 
The corresponding step heights vary within an order of $1 \%$ at the beginning of a dynamical process, so that those in Fig. \ref{FIG:1}(c) (except for the first steps) look rather uniform. The evolution time is counted by the oscillation period number $t/T\sim \omega_m t$. (b) The frequency comb of the IPP pulse due to $\omega_m/\kappa=0.1$. We here fix $g_m/\kappa=10^{-5}$ and $\gamma_m/\kappa=10^{-5}$, and use two drives with $E/\kappa=10^5$ and $\omega_1=\omega_c$, $\omega_2=\omega_c-\omega_m$.}
	\label{FIG:3}
\end{figure}

\section{Dynamical pattern in resonance}
\subsection{Relevant phenomena}
A particular choice for the setup in Fig. \ref{FIG:1}(a) is that the two driving fields with the same amplitude $E$ keep their frequency difference as $|\omega_1-\omega_2|=\omega_m$. Under this condition, a resonance phenomenon manifests to have the mechanical energy $\mathcal{E}_m(t)=1/2\left(X_m^2(t)+P_m^2(t)\right)$ evolving as the temporal stairs in Fig.\ref{FIG:1}(c). While the length of their steps is the same as the corresponding mechanical oscillation period $T=2\pi/\omega_m$, a temporal stair for each fixed $\omega_m$ becomes steeper with the nonlinear magnitude $g_m$ enhanced by the drive amplitude $E$ [see Fig. \ref{FIG:1}(d)]. The energy transfer is like a repeated quanta $\hbar\omega_m \delta\mathcal{E}_m$ ($\delta\mathcal{E}_m$ is the step height equivalent to a phonon number) from the cavity field to the mechanical oscillator in each period $T$. Viewed as a separate part, the mechanical motion takes the form
\begin{eqnarray}
P_m(t)&=&g_m\int_0^t d\tau e^{-\gamma_m (t-\tau)}\{\cos\frac{1}{2}\sqrt{4\omega_m^2-\gamma_m^2}(t-\tau)\nonumber\\
&-&\frac{\gamma_m\sin\sqrt{4\omega_m^2-\gamma_m^2}(t-\tau)}{\sqrt{(4\omega_m^2-\gamma_m^2)}}\}|a(\tau)|^2
\label{sol}
\end{eqnarray}
from the last two equations in Eq. (\ref{2}), necessitating a pulsed field intensity $|a(t)|^2$ for the step-like change 
of the energy $\mathcal{E}_m(t)$. Pulsed cavity field due to pumping unresolved-sideband OMSs ($\omega_m<\kappa$) by a single-frequency field was theoretically predicted \cite{poot2012backaction,miri2018optomechanical,xu2021chip} and experimentally observed \cite{hu2021generation}. However, under two drives satisfying the condition $|\omega_1-\omega_2|=\omega_m$, why there exists a particular evolution pattern demonstrating the temporal stairs with their uniform steps in Fig. \ref{FIG:1}(c) should be better understood from comparing two scenarios simulated in Fig. \ref{FIG:2}.

The first scenario, with $|\omega_1-\omega_2|$ mismatching $\omega_m$ by $\delta=0.05\kappa$ ($|\omega_1-\omega_2|=\omega_m+\delta$), 
gives the pulsed cavity field intensities $|a(t)|^2$ around each half mechanical oscillation period $t=nT/2$ ($n$ is an integer) as in Fig. \ref{FIG:2}(a). These pulses provide the in-phase pushes (IPP) when they act along the same direction of the moving oscillator (around each positive maximum $P_m$), but work as the out-of-phase pushes (OPP) when they oppose the mechanical motion around each negative maximum $P_m$. The real-time energy $\mathcal{E}_m(t)$ thus undergoes alternate jumps because the pulse height changes with its period $(\omega_m/\delta)T=4T$ in this situation. There are other variation patterns of the mechanical energy and the pulses off the resonance 
point $|\omega_1-\omega_2|=\omega_m$, as we will see in Sec. 4.1. To the second resonance scenario ($\delta=0$) in Fig. \ref{FIG:2}(b), the OPP pulses are highly suppressed but the IPP pulses seem to stabilize gradually as in Fig. \ref{FIG:3}(a). The IPP pulses sustain the repeated energy addition $\hbar\omega_m \delta\mathcal{E}_m$ to the oscillator until its displacement $X_m$ goes beyond the valid range for the potential in Eq. (\ref{potential}), and their Fourier transform in Fig. \ref{FIG:3}(b) shows an optical frequency comb (OFC) with its teeth distanced by the mechanical frequency $\omega_m$.  

\begin{figure}
	\centering
		\includegraphics[scale=.058]{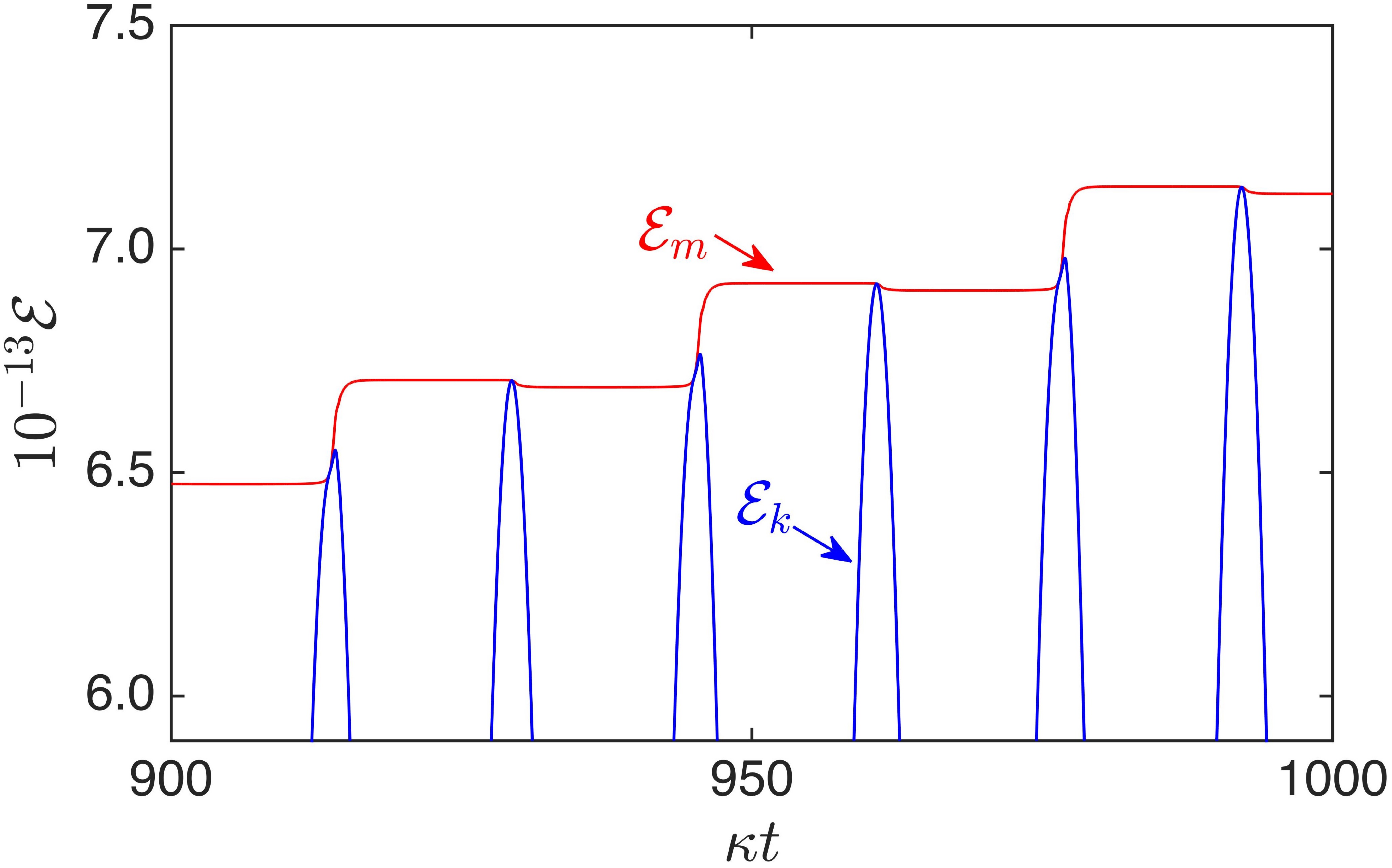}
	\caption{{\bf Comparison between the total mechanical energy and its kinetic part.} The kinetic energy $\mathcal{E}_k(t)=\frac{1}{2}P_m^2(t)$ reaches its maximum whenever the mechanical oscillation is at each half cycle $t=nT/2$ ($n$ is an integer), along with the evolution of the total mechanical energy $\mathcal{E}_m(t)=\frac{1}{2}\left(X_m^2(t)+P_m^2(t)\right)$. In the middle of each step for the mechanical energy $\mathcal{E}_m(t)$, the action of the pulsed field is in the opposite direction of the mechanical motion, while the pulsed field speeds up the mechanical oscillator around each step jump of the mechanical energy. This comparison clearly shows that the dimensionless momentum $P_m(t)$ around each step jump of the energy $\mathcal{E}_m(t)$ is higher than the $P_m(t)$ around a previous one. Here the system parameters are the same as those in Fig. \ref{FIG:2}(b).}
	\label{FIG:4}
\end{figure}

\begin{figure}
	\centering
		\includegraphics[scale=.07]{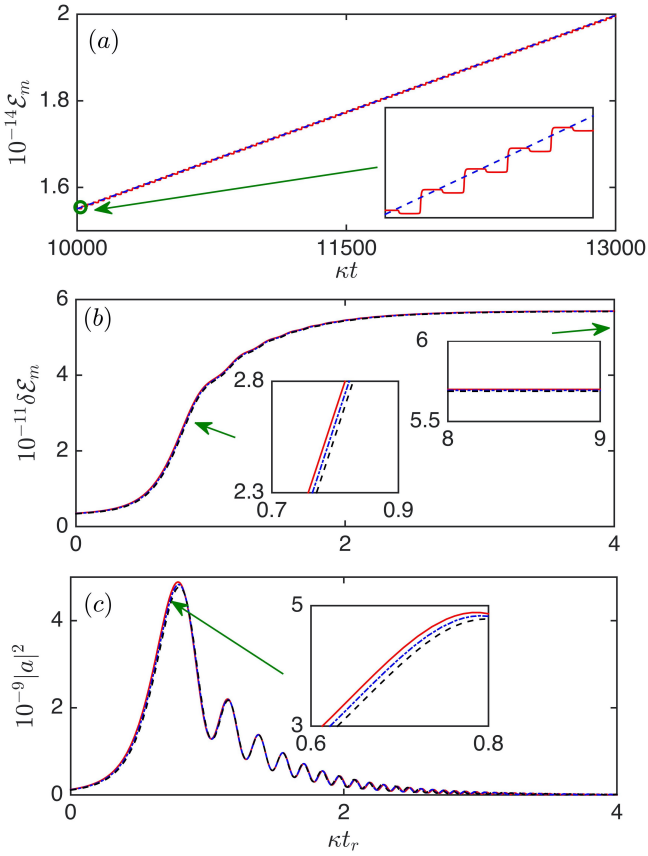}
	\caption{{\bf A self-organized cooperation between two subsystems.} (a) The fitting of the mechanical energy $\mathcal{E}_m(t)$ has a completely linear tendency over a sample duration $\kappa\Delta t=3000$ from the moment $\kappa t=10000$ to another moment $\kappa t=13000$. (b) and (c) The details of three steps, around $\kappa t=10100$ (red solid), $\kappa t=10100+100\pi$ (blue dot-dashed), and $\kappa t=10100+200\pi$ (black dashed), extracted along the fitted straight line in part (a), together with the associated cavity field pulses that create these steps. The steps and their corresponding pulsed field intensities $|a|^2$ are all displaced to the common starting moment of a reference time $t_r$ for the purpose of their comparison. The insets show the details in the developments of the temporal steps and the corresponding field intensities $|a|^2$. Each stabilized part of the steps [beyond Fig. \ref{FIG:5}(b)] completely coincides to realize the same step height and the linear tendency in (a). The exemplary OMS under the drives with the same amplitude $E/\kappa=10^5$ has the same systems parameters as those in Fig. \ref{FIG:2}.}
	\label{FIG:5}
\end{figure}

\subsection{Example of field-oscillator cooperation}
In terms of the model in Fig. \ref{FIG:1}(b), the second scenario in Fig. \ref{FIG:2} is that the oscillator modeling the cavity field is excited around each moment $t=nT$ and, after the duration of the pulsed $|a(t)|^2$, it drops back to near the bottom of the binding potential. The excited one happens to bring an impulse $I_p$ to the other one around its highest positive speed at $X_m(nT)=0$, so this mechanical oscillator will go faster whenever it cycles back to the point $X_m((n+1)T)=0$ in the next oscillation period (for an oscillator with high quality factor the energy loss in each oscillation period is much less than the energy added by the pulsed field). This fact can be seen more clearly from Fig. \ref{FIG:4} showing the comparison between the dimensionless kinetic energy $\mathcal{E}_k(t)=1/2P_m^2(t)$ of the oscillator and its whole mechanical energy $\mathcal{E}_m(t)=1/2\left(X_m^2(t)+P_m^2(t)\right)$. Moreover, the momentum $P_m(t)$ of the mechanical oscillator keeps a smooth evolution during its collision with the pulsed field; compare the evolution course of $P_m(t)$ with that of $\mathcal{E}_m(t)$ 
in Fig. \ref{FIG:2}(b). 

The realized interaction under the condition $|\Delta_1-\Delta_2|=\omega_m$ for two driving fields brings about a perfect linear tendency for the evolution of the mechanical energy $\mathcal{E}_m(t)$; the completely straight line [the dashed one shown in the inset of Fig. \ref{FIG:5}(a)] fitted for the energy $\mathcal{E}_m(t)$ lasts over a duration $\kappa \Delta t=3000$ for the example in Fig. \ref{FIG:5}(a). It means that the mechanical energy $\mathcal{E}_m(t)$ keeps jumping along the straight stair by each constant step height $\delta\mathcal{E}_m$ around the moments $nT$ ($n$ is an integer).
Suppose that the $n$-th tall pulse in Fig. \ref{FIG:2}(b) acts within a small time window $t\in [nT-\delta t_{n,1},nT+\delta t_{n,2}]$ near the moment $nT$ (the $n$-th oscillation cycle from the beginning of the time evolution), to have its whole duration $\delta t_n=\delta t_{n,1}+\delta t_{n,2}$. Multiplying $dX_m(t)$ on both side of the third equation in Eq. (\ref{2}) and performing an integral over the pulse duration $\delta t_n$, one will have 
\begin{eqnarray}
\delta\mathcal{E}_m&=&\frac{1}{2}\left(P^2_m(nT+\delta t_{n,2})+X^2_m(nT+\delta t_{n,2})\right)\nonumber\\
&-&\frac{1}{2}\left(P^2_m(nT-\delta t_{n,1})+X^2_m(nT-\delta t_{n,1})\right)\nonumber\\
&=&g_m\int_{nT-\delta t_{n,2}}^{nT+\delta t_{n,1}}dt |a(t)|^2P_m(t)
\label{work}
\end{eqnarray}
as the step height of the mechanical energy $\mathcal{E}_m(t)$ around the moment $t=nT$. 
Considering the fact $\gamma_m\ll \omega_m$, we here neglect the contribution from the mechanical damping force during the small time window $\delta t_n$. The momentum $P_m(t)$ after one more oscillation period (the $P_m(t)$ for $t\in [(n+1)T-\delta t_{n+1,1},(n+1)T+\delta t_{n+1,2}]$) is higher than the $P_m(t)$ for $t\in [nT-\delta t_{n,1},nT+\delta t_{n,2}]$, as seen from Fig. \ref{FIG:4}. To have the same definite integral around these two moments $t=nT$ and $t=(n+1)T$ according to Eq. (\ref{work}) so that the mechanical energy jumps around the two moments have the same height, the pulsed field intensity $|a(t)|^2$ should adjust itself following the increased momentum $P_m(t)$ around the next step jump of the energy $\mathcal{E}_m(t)$. Then it will build up a kind of their synchronization within the range like the one in Fig. \ref{FIG:5}(a). The effect of such synchronization is that the pulsed field lasting for each short period $\delta t_n$ ($n$ as a series of integers) happens to do the exactly same amount of work ($\delta\mathcal{E}_m=\text{constant}$) to the mechanical oscillator, which runs faster and faster. 

\begin{figure}
	\centering
		\includegraphics[scale=.041]{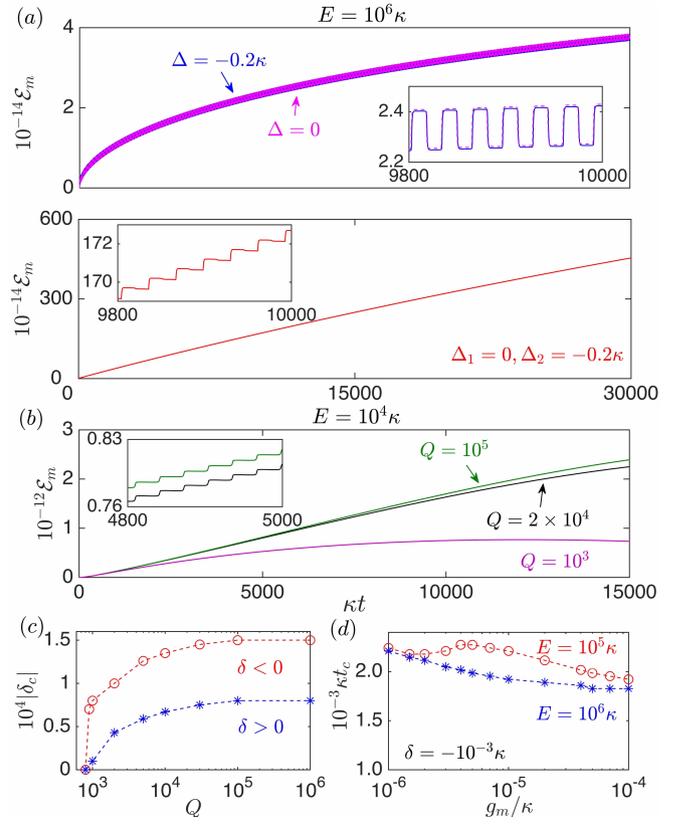}
	\caption{{\bf Resonance and its breakdown.} (a) A comparison between the single-drive scenarios 
and the resonance under two simultaneous drives satisfying $|\omega_1-\omega_2|=\omega_m$. The nearly identical $\mathcal{E}_m(t)$ for both individual drives with their different $\Delta$ is bound as the illustrated asymptotic stability. We adopt the definition $\Delta=\omega_c-\omega$, and the system has $\omega_m/\kappa=0.2$, $g_m/\kappa=10^{-5}$, and $\gamma_m/\kappa=10^{-5}$. (b) The corresponding evolution courses of the energy $\mathcal{E}_m(t)$ after $\Delta_1\rightarrow \Delta_1$, $\Delta_2\rightarrow \Delta_2+\delta$, with $\delta=-10^{-4}\kappa$. The lower quality factors $Q=\omega_m/\gamma_m$ make the courses deviate from the linear tendency more obviously. (c) The allowed maximum error $\delta_c$ to have the temporal stairs extending over $\kappa t=1.5\times 10^4$. Here $E/\kappa=10^4$. (d) The maximum duration $\kappa t_c$ of the temporal stairs after an existing error $\delta$ from the drive-frequency condition. Decreasing or increasing $g_m$ does not change the tendency so much.}
	\label{FIG:6}
\end{figure}

To see how a pulsed field force precisely add a constant amount of energy or do the same amount of work during each small time window $\delta t_n$ within the illustrated range, we pick out three equally distanced steps along the fitted straight line on the temporal stair in Fig. \ref{FIG:5}(a), and displace these steps and the associated pulses that create them to a common staring point to compare their respective dynamical evolution process in Figs. \ref{FIG:5}(b) and \ref{FIG:5}(c). From one sample process to another, one finds that the pulses shown in Fig. \ref{FIG:5}(c) are fine-tuning their fronts by themselves so that the rise of the peak for a pulse that pushes the faster oscillator slightly delays by a proper amount; the details are illustrated in the inset of Fig. \ref{FIG:5}(c). Due to such a delicate adjustment, the different energy additions to the mechanical oscillator, as the heights of the temporal steps, happen to be the same---the three evolved horizontal lines in the upper right inset of Fig. \ref{FIG:5}(b) completely overlap. For the example in Fig. \ref{FIG:2}(b), the step height $\delta\mathcal{E}_m=5.623\times 10^{11}$ near $\kappa t=10^3$ only lowers to $\delta\mathcal{E}_m=5.183\times 10^{11}$ till $\kappa t=10^5$, as the self-adjustment of the pulse action also modifies over a larger time scale. Confined to the straight-line increase of mechanical energy $\mathcal{E}_m(t)$ in Fig. \ref{FIG:5}(a), the coupled pulses and mechanical oscillator reach a perfect mutual coordination or are locked together into a unique pattern of energy transfer. This could be viewed as a generalization of the phase synchronization \cite{pikovsky2002synchronization, acebron2005kuramoto,boccaletti2018synchronization} between two harmonic oscillations---two totally different types of motion (one is harmonic oscillation but the other is pulsed excitation) are self-organized into a unique way of energy transfer.  
In nonlinear dynamical processes it is common to see the locking phenomena in a parameter space (the well-known Devil's staircase) or a locked step-by-step evolution of oscillation phase (see, e.g. \cite{pikovsky2002synchronization}), but a perfect cooperation between two subsystems to realize a constantly and uniformly step-like energy addition to one of them is beyond the previously studied phenomena of synchronization \cite{pikovsky2002synchronization,boccaletti2018synchronization}.

\subsection{Pattern locking mechanism}

Driven by a single pump, the responding mechanical motion never reaches a resonance. Unlike in the regime $\omega_m/\kappa>1$, the induced mechanical motion in the regime $\omega_m/\kappa<1$ can be nearly identical when scanning the single-drive frequency from a blue-detuned to a red-detuned; see one example in Fig. \ref{FIG:6}(a). In contrast, two simultaneous drives lead to a boundless energy harvesting at the point $|\omega_1-\omega_2|=\omega_m$, due to a triggered mechanism to lock the system into a cooperative energy transfer. This mechanism even works under the drives of very low power $E/\kappa\ll 1$ (see Sec. 4.2), though the associated cavity fields are not pulses. A Hopf bifurcation from a steady to an oscillating OMS occurs when the amplitude $E$ of a single drive should be over a threshold for the OMSs in the regime $\omega_m/\kappa>1$, but the above-mentioned mechanism brings an unresolved-sideband OMS to oscillation without a restriction on drive power. The dynamical pattern of resonance cannot be preserved and gradually turns into something like the one in Fig. \ref{FIG:2}(a) if $|\omega_1-\omega_2|$ keeps going away from $\omega_m$, but it can be compensated to some extent by a higher mechanical quality factor, as shown from Figs. \ref{FIG:6}(b) and \ref{FIG:6}(c). In the ideal situation $\gamma_m=0$, the perfect stairs of $\mathcal{E}_m(t)$ last longer than $\kappa t=10^4$ in spite of an error $|\delta|=10^{-4}\kappa$ from the condition $|\omega_1-\omega_2|=\omega_m$, under which the linearly rising tendency of $\mathcal{E}_m(t)$ also disappears due to a larger damping rate $\gamma_m$ (see Sec. 4.3). Fig. \ref{FIG:6}(d) shows that changing the magnitude $g_m$ itself cannot improve the process. All these facts further prove that the mechanism of pattern locking is rather different from all other types of nonlinearity; a mismatch between $|\omega_1-\omega_2|$ and $\omega_m$ or more energy loss will finally impair the cooperation between two subsystems. 

\section{Scenarios beyond perfect resonance}

\subsection{Dynamical patterns off resonance point}
\begin{figure*}
	\centering
		\includegraphics[scale=.34]{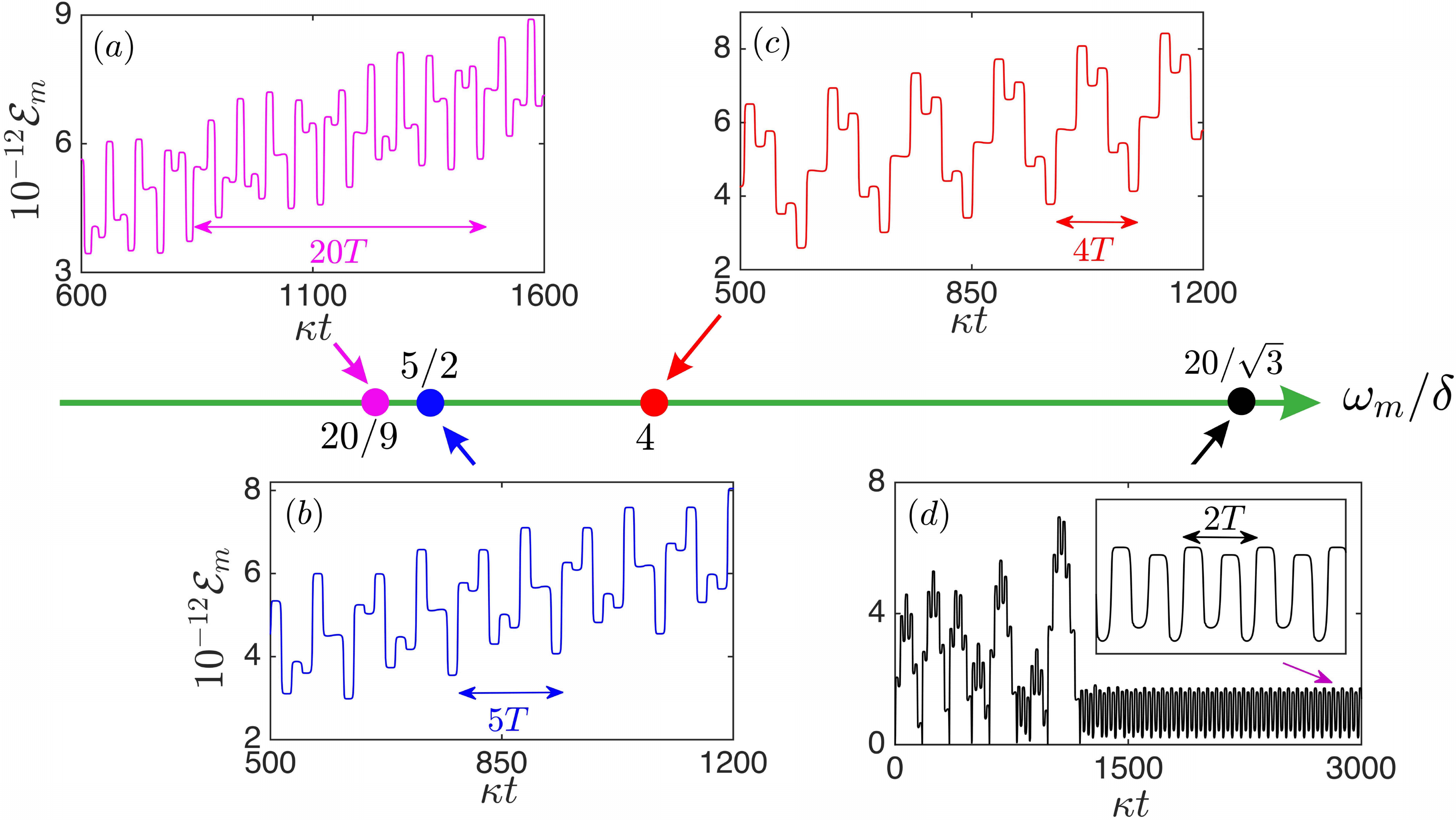}
	\caption{{\bf Periodic mechanical energy patterns off the resonance point,} The stabilized period is determined by the ratio $\omega_m/|\delta|$. The evolution in (d) with an irrational ratio $\omega_m/\delta$ first undergoes an aperiodic stage before its stabilization to a periodic pattern. The setup in these processes has the same parameters as those in Fig. \ref{FIG:2} and $E/\kappa=2\times 10^5$. }
	\label{FIG:7}
\end{figure*}

In practice it is much easier to have two pumping tones off the resonant point, i.e. $|\omega_1-\omega_2|=\omega_m+\delta$ with an error $\delta$. If the error is tiny, such as $|\delta|=10^{-4}\kappa- 10^{-3}\kappa$ for the examples in Fig. \ref{FIG:6}, the mechanical energy evolution will become a 
ladder that gradually bends with time. At somewhere further away from the resonance point, there will appear rich varieties of local patterns for the energy $\mathcal{E}_m(t)$ and its associated cavity field pulses, in addition to the overall asymptotic stability for the real-time $\mathcal{E}_m(t)$ (the system stabilizes at $t\rightarrow\infty$). In Fig. \ref{FIG:7} we only display some patterns for the jumps of the energy $\mathcal{E}_m(t)$, since the associated change of the field pulse height can be deduced with their one-to-one correspondence as in Fig. \ref{FIG:2}. 

\begin{figure}
	\centering
		\includegraphics[scale=.049]{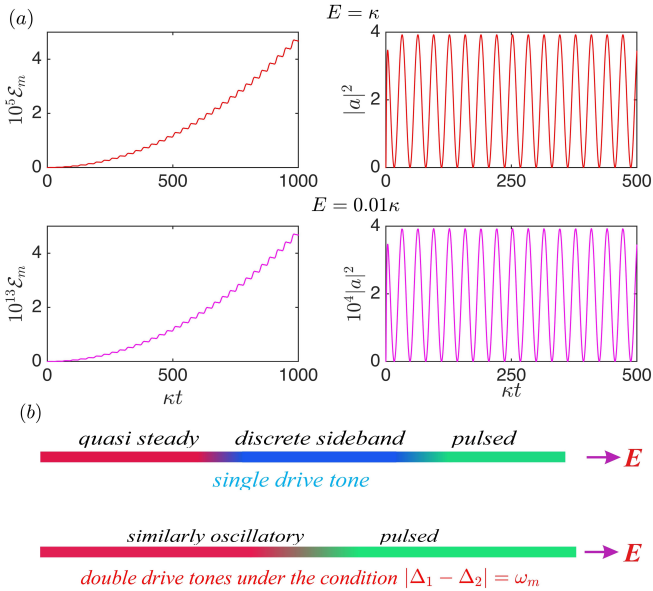}
	\caption{{\bf Scenarios under arbitrary drive power.} (a) The pattern of similar oscillation for the systems under low drive powers but still meeting the resonance condition. The system parameters for the exemplary OMS are the same as those in Fig. \ref{FIG:2}. (b) The comparison between the effects of drive power for a single tone and a double-tone drive under the frequency condition for realizing the mentioned resonance. Three patterns exist in the former, while the latter has only two patterns that gradually take transition to each other by tuning the drive amplitude $E$.  }
	\label{FIG:8}
 \vspace{-0.3cm}
\end{figure}

We find that such dynamical patterns off the resonance point are completely determined by the ratio $\omega_m/|\delta|$. Given a rational ratio $\omega_m/\delta=m/n$, where $m$ and $n$ are two relatively prime integers, the period for a round of variation of the pulse height, which displays the patterns illustrated in Fig. \ref{FIG:7}, is $N(\omega_m/\delta)T$ with $N$ being an integer and $T=2\pi/\omega_m$ being the mechanical oscillation period. For example, the variation period will be $T$, the pattern of the single drives in Fig. \ref{FIG:6}(a), if the error $\delta$ is 
up to $-\omega_m$. The exemplary patterns in Figs. \ref{FIG:7}(a), \ref{FIG:7}(b) and \ref{FIG:7}(c) have their periods $9(\omega_m/\delta)T$, $2(\omega_m/\delta)T$ and $(\omega_m/\delta)T$, respectively. More interesting is an irrational ratio $\omega_m/\delta$ as in Fig. \ref{FIG:7}(d). In this situation the system first undergoes an aperiodically transient stage and then stabilizes to a particular periodic pattern, but the stabilized period does not follow a fixed law. Because the rational numbers are like isolated islands in the ocean of the irrational numbers, the period for the dynamical patterns changes randomly along the axis $\omega_m/\delta$ in Fig. \ref{FIG:7}. Different errors $\delta$ make the system be locked into different patterns of periodic variation for the pulse height and the mechanical energy jumps. The resonance point at $\delta=0$ can be regarded to have an infinitely long period with $\omega_m/\delta\rightarrow \infty$.

\subsection{Mechanical resonance under extremely low drive power}

The resonance phenomenon under the condition $|\Delta_1-\Delta_2|=\omega_m$ exists even when the drive powers are very low, as shown in Fig. \ref{FIG:8}(a). In fact, this illustrated scenario of the similar responses of the system to the varied drive amplitude $E$ even exists when $E\rightarrow 0$, although the increase of the mechanical energy $\mathcal{E}_m(t)$ loses the linear tendency even at the beginning stage. A remnant cooperation between two subsystems, due to the action of a mechanism to lock them into the pattern, still helps to push the mechanical oscillation to higher amplitude until after a relatively long duration. 

On the other hand, when the same system is driven by a single-tone field, there exists a gradual transition from a quasi-steady (like the observation reported in Ref. \cite{doolin2014nonlinear}) to an oscillating field once the drive amplitude is over the level $E/\kappa\sim10^3$ for the used OMS. If the pump power of the single drive is increased further, the systems in the regime $\omega_m/\kappa<1$ will respond almost identically to the differently detuned driving fields [see Fig. \ref{FIG:6}(a)]. In Fig. \ref{FIG:8} (b) we summarize the dynamical patterns due to a varied drive amplitude $E$ in both scenarios of a single-tone and a double-tone drive meeting the condition $|\Delta_1-\Delta_2|=\omega_m$. To both of the scenarios, there is no bifurcation on the way of tuning the drive power because, unlike near a Hopf bifurcation in the regime $\omega_m/\kappa>1$, the gradual transitions between different patterns do not show a phenomenon of critical slowing-down. 

\begin{figure}
	\centering
		\includegraphics[scale=.055]{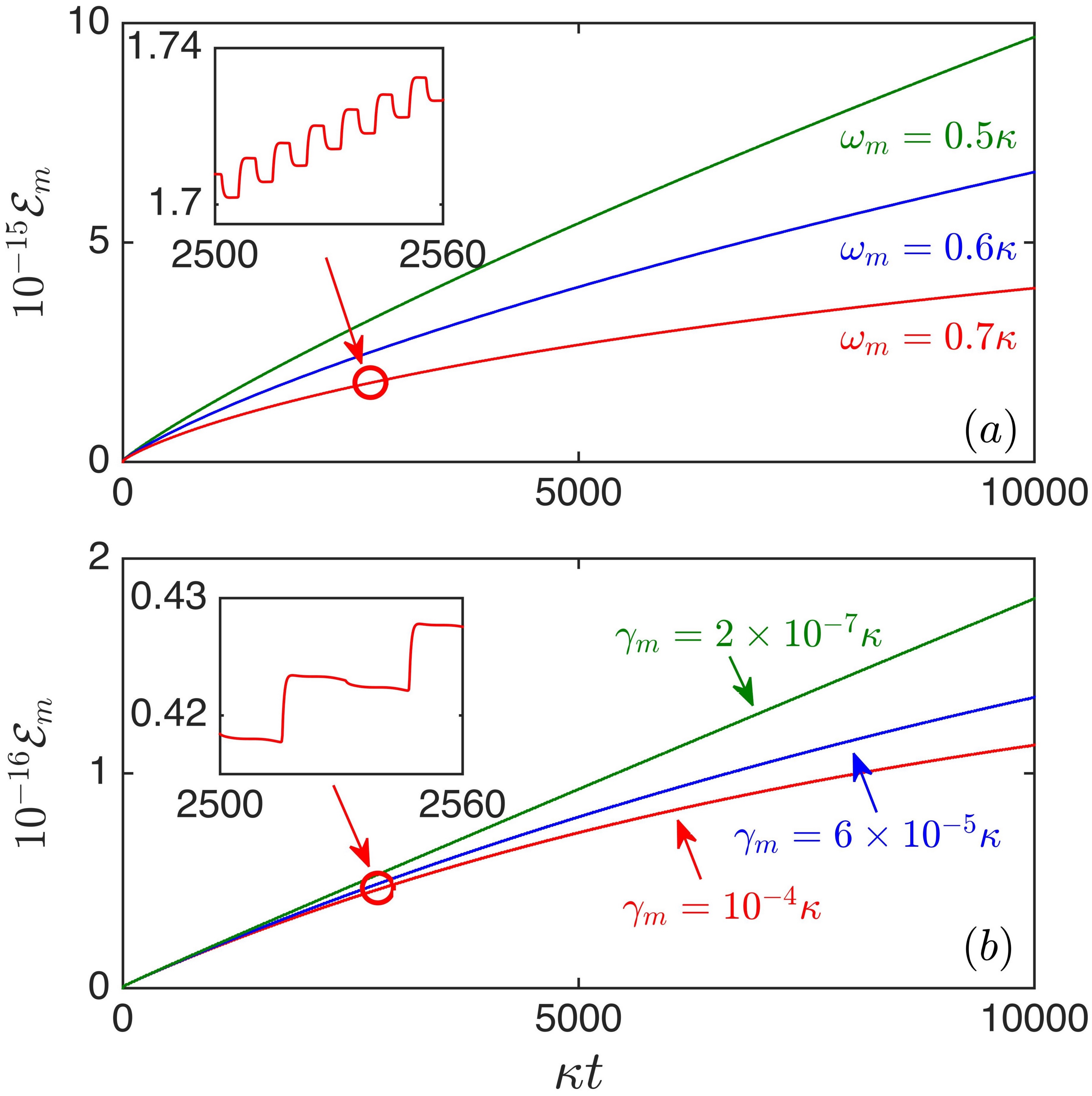}
	\caption{{\bf Factors that damage straight stair-like tendency.} (a) The evolution processes for the different $\omega_m$, with a vanishing mechanical damping $\gamma_m=0$ for each $\omega_m$. (b) The evolution processes 
for the different $\gamma_m$, but with a fixed mechanical frequency $\omega_m=0.2\kappa$. Here, we consider an OMS with $g_m=10^{-5}\kappa$ and apply two driving fields with the same amplitude $E=10^6\kappa$ and their detunings $\Delta_1=0$ and $\Delta_2=\omega_m$. }
	\label{FIG:9}
\end{figure}

\subsection{Stronger spring force and more energy loss}

Changing two other parameters can destroy the linearly rising stairs for the energy $\mathcal{E}_m(t)$, even if the system is still under the resonance condition. 
In Fig. \ref{FIG:9}(a) we display a group of sample evolution processes with their only difference in 
the mechanical frequency $\omega_m$. Here, we assume an ideal situation $\gamma_m=0$ (the dissipation 
in the system is only through the cavity field damping $\kappa\neq 0$), so the dynamics is simplified to 
see the essential effect due to the mechanical frequency change. From the numerical simulations of the pulsed 
$|a|^2$ shown in the inset of Fig. \ref{FIG:3}(a), one sees that, in the regime $\omega_m<\kappa$, the peak heights of the IPP pulses go down with the mechanical frequency $\omega_m$ but those of the OPP pluses grows up with $\omega_m$. Then the evolving $\mathcal{E}_m(t)$ for a higher $\omega_m$ will become square-shaped [see the inset in Fig. \ref{FIG:8}(a)]. Given a higher $\omega_m$, the action of the IPP pulses also becomes less matched to the mechanical oscillation around its positive maximum speed, thus losing the dynamical pattern of the cooperation between two subsystems. Another factor that is illustrated in Fig. \ref{FIG:9}(b) is the mechanical damping rate $\gamma_m$, which causes the loss 
\begin{eqnarray}
\Delta\mathcal{E}_m=-\gamma_m\int_{nT}^{(n+1)T} P_m^2(t)dt
\end{eqnarray}
of the mechanical energy
$\mathcal{E}_m(t)=1/2\left(X_m^2(t)+P_m^2(t)\right)$ in each oscillation period $T$. At a larger mechanical oscillation amplitude, more energy loss will be experienced by the oscillator that is running faster to have higher $P^2_m$. The input energy from the pulsed field will not be sufficient to keep the linear increase of the energy $\mathcal{E}_m(t)$. 

\begin{figure}
	\centering
		\includegraphics[scale=.055]{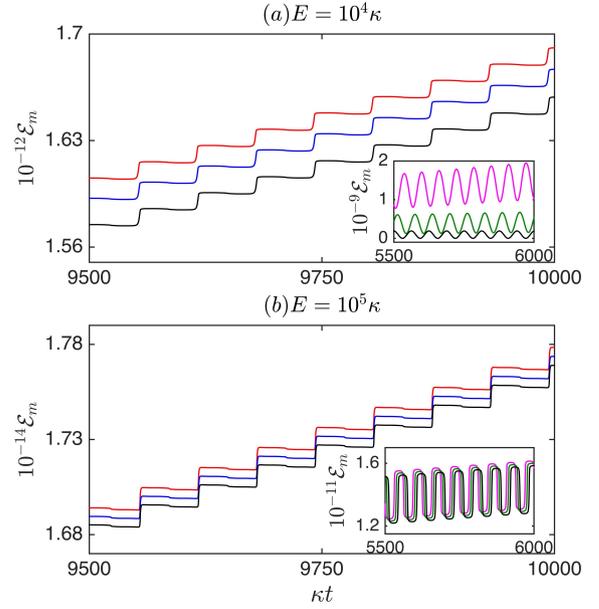}
	\caption{{\bf Resonance under parallel drive-frequency shifts.} (a) The examples of the created mechanical energy $\mathcal{E}_m$ by the three combinations $(\Delta_1,\Delta_2)=(-0.1\kappa,0)$ (red), $(\Delta_1,\Delta_2)=(-0.05\kappa,0.05\kappa)$ (blue), and $(\Delta_1,\Delta_2)=(0,0.1\kappa)$ (black), which all have the difference $|\Delta_1-\Delta_2|$ matching the used mechanical frequency $\omega_m=0.1\kappa$. By employing the single drives, $\Delta=-\omega_m$ (pink), $\Delta=0$ (green), and $\Delta=\omega_m$ (dark), respectively, the realized mechanical energy will be those shown in the inset. Here, the drive amplitude is $E=10^4\kappa$. (b) The corresponding result by increasing the drive amplitude to $E=10^5\kappa$. The system has $Q=\omega_m/\gamma_m=10^4$ and $g_m/\kappa=10^{-5}$. }
	\label{FIG:10}
\end{figure}

 \section{Discussions on other related issues}
 
\subsection{Parallel shift of drive frequencies}
Previously, OMSs driven by multiple driving fields were studied theoretically (see, e.g. \cite{mari2009gently}) and experimentally \cite{shomroni2019two} in the regime where $\omega_m/\kappa>1$. A more interesting scenario that is predicted for the regime is locking the mechanical oscillation amplitude to some fixed values under the condition $|\omega_1-\omega_2|=\omega_m$, in spite of varying the drive power; this scenario realizes a series of fixed stable 
mechanical orbits like energy levels
\cite{he2020mechanical,wu2022amplitude,li2022ultra,yan2023force}. These frozen mechanical orbits still exist after shifting the frequencies of two driving fields together, given a possible adjustment of the drive power. 

Our concerned resonance phenomenon in the regime of 
$\omega_m/\kappa<1$ is also preserved under the parallel shift 
\begin{eqnarray}
&&\omega_1\rightarrow \omega_1+\delta,\nonumber\\
&&\omega_2\rightarrow \omega_2+\delta
\label{shift}
\end{eqnarray}
of the frequencies of two driving fields.
Once the absolute difference $|\omega_1-\omega_2|$ is close to the mechanical frequency $\omega_m$, a suitable system pumped by two drives that are not too weak will demonstrate the resonance phenomenon. In Fig. \ref{FIG:10} we compare the temporal steps of the energy $\mathcal{E}_m(t)$ that are realized by the different combinations of $\Delta_1$ and $\Delta_2$ under the condition $|\Delta_1-\Delta_2|=\omega_m$. These temporal stairs under such condition are parallel to one another, but the most blued-shifted combination leads to the highest mechanical energy. To maintain a fixed distance $|\omega_1-\omega_2|$ is, therefore, a technical prerequisite to realize the dynamical scenario. A similar condition was required in a previous experiment performed in a resolved sideband regime ($\omega_m/\kappa>10$) \cite{shomroni2019two}, where two drive tones are tuned to near $\omega_c\pm\omega_m$ to see a type of dynamical instability.   

As a comparison, an individual action of the involved single driving fields can give rise to different responses of the mechanical oscillator. The examples are given in the inset of Fig. \ref{FIG:10}(a). By increasing the pump power, the evolution trajectories of $\mathcal{E}_m(t)$ will become more and more identical, as seen from those in the inset of Fig. \ref{FIG:10}(b) to those in an inset of Fig. \ref{FIG:6}(a). The contrast between the doubly driven and the singly driven scenario in Fig. \ref{FIG:10}(a) is dramatic. Under that used lower pump amplitude $E=10^4\kappa$ for the exemplary system, any single-tone drive can never create a pulsed cavity field that realizes the square-shaped $\mathcal{E}_m(t)$ shown in the inset of Fig. \ref{FIG:10}(b). However, a combination of two tones simply satisfying the condition $|\Delta_1-\Delta_2|=\omega_m$ will generate a pulsed field that drives the mechanical oscillator to repeated sudden jumps of its energy. This is another illustration of enhancing the nonlinearity under the drive-frequency condition $|\Delta_1-\Delta_2|=\omega_m$. 

\begin{figure}
	\centering
		\includegraphics[scale=.063]{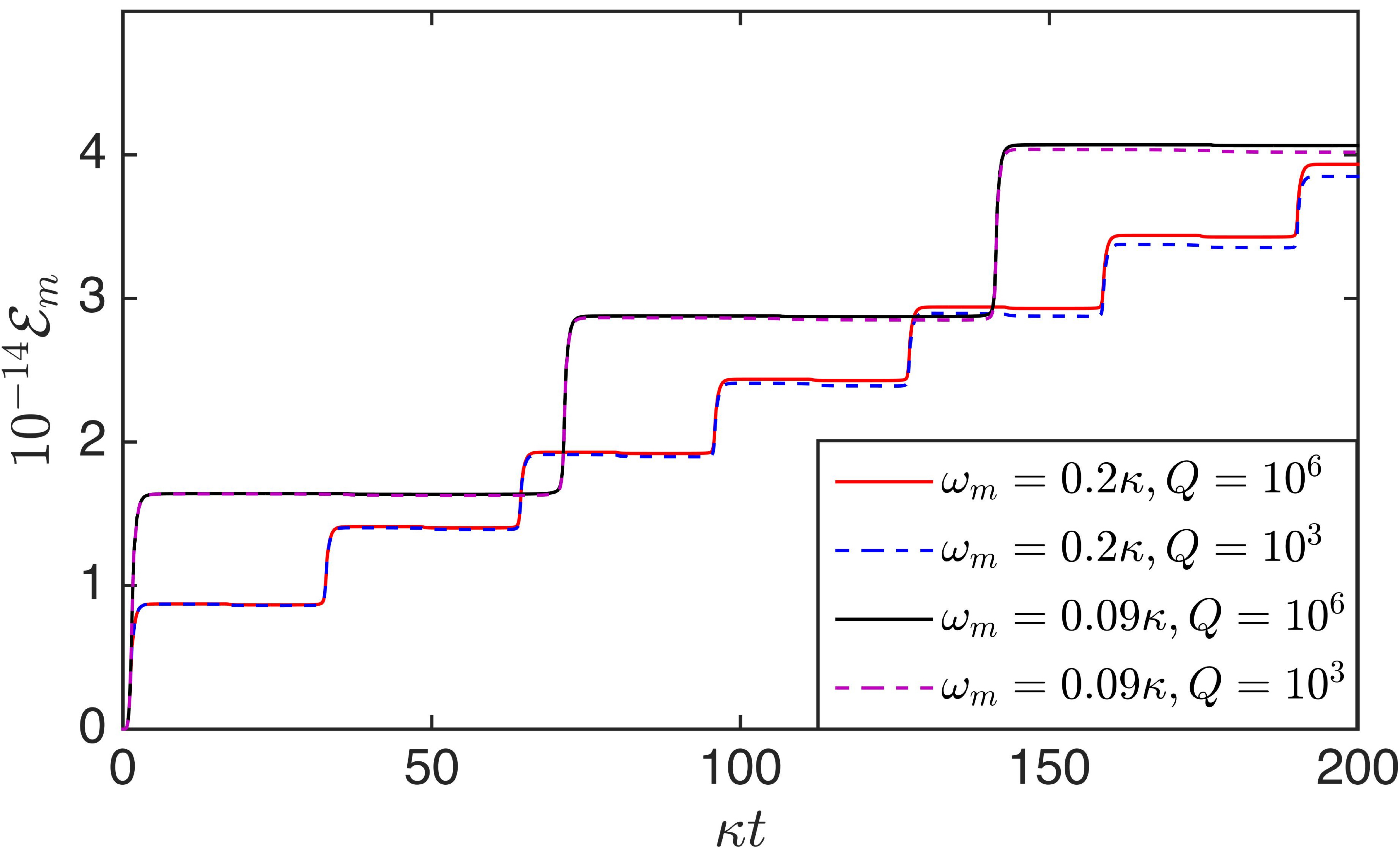}
	\caption{{\bf Insignificant influence of thermal noise.} Compared with the results at zero temperature, the thermal perturbation at the room temperature does not bring about much influence on the temporal stairs of the mechanical energy. The fixed system parameters for the exemplary systems are the same as those in Fig. \ref{FIG:2}, and the specific cavity damping rate is $\kappa=2\pi\times 30$ MHz. }
	\label{FIG:11}
\end{figure}

\subsection{Thermal noise effect}
The dynamical scenarios related to our concerned resonance are characterized by gradual transitions, instead of sudden transitions of bifurcation in many other nonlinear systems. For example, the loss of the linearly increasing tendency of $\mathcal{E}_m(t)$, as the bent 
temporal stairs displayed in Fig. \ref{FIG:9}, is gradual. 
One consequence of this feature is in the numerical integral of Eq. (\ref{2}); all of our numerical calculations are not obviously affected by the adopted precision, 
in contrast to those near a bifurcation point where the critical dynamics makes the simulation results highly sensitive to the used computation precision 
(see, e.g. Supplemental Material of \cite{lin2021catastrophic}). 

Noise perturbations are also less important to the concerned dynamical process. To study a specific example, we adopt a thermal noise term in Eq. (\ref{2}). The thermal noise at room temperature can be approximated by a white noise with the correlation $\langle \xi_m(t)\xi_m(t')\rangle\approx 2n_{th}\delta(t-t')$ ($n_{th}$ is the thermal occupation at a certain temperature) \cite{lin2020entangling}. This noise term can be simulated by a random function \cite{lin2021catastrophic}. In Fig. \ref{FIG:11} we display the evolving stairs of the mechanical energy $\mathcal{E}_m(t)$ at the room temperature $300$ K for two exemplary $\omega_m$. One sees that the influence of the environmental temperature is negligible. If we reduce the mechanical quality factor $Q=\omega_m/\gamma_m$ by $10^3$ times, the realized step heights will be slightly lowered. The dynamical pattern with the cooperative energy transfer between cavity field excitation and mechanical oscillation is rather robust against thermal perturbation. 

\begin{figure}
	\centering
		\includegraphics[scale=.045]{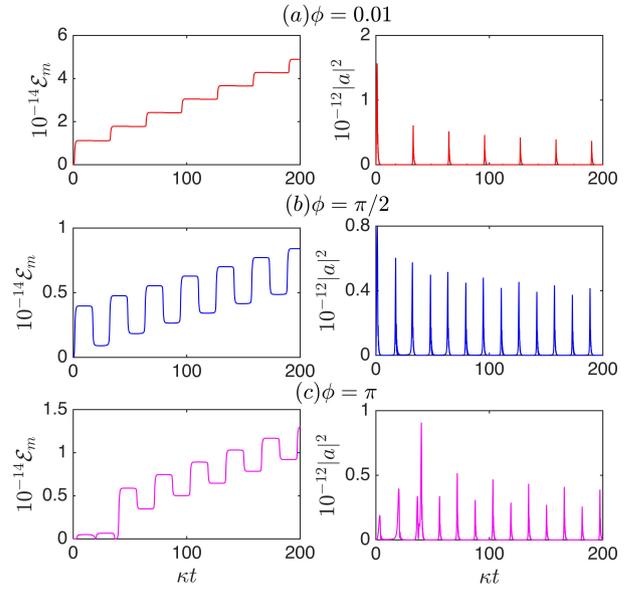}
	\caption{{\bf Effect of unmatched driving field phases.} (a) A small deviation $\phi=0.01$ can hardly impair the dynamical pattern that realizes the repeatedly uniform steps of the mechanical energy. (b) and (c) 
The nonlinear interference results due to the large relative phases $\phi$. Different patterns of the pulsed field, corresponding to the illustrated  mechanical energy, can be created by adjusting the phase. The system parameters for the exemplary OMS are the same as those in Fig. \ref{FIG:2}.}
	\label{FIG:12}
\end{figure}

\subsection{Phase mismatch between driving fields}

The two fields as the driving terms in Eq. (\ref{2}) may have different phases $\theta_1$ and $\theta_2$ in reality. We here illustrate the effect of the relative phase $\phi=|\theta_1-\theta_2|$ by the numerically calculated $\mathcal{E}_m(t)$ and $|a(t)|^2$ in Fig. \ref{FIG:12}. Deviating the phase $\phi$ from the ideal condition $\phi=0$, one will obtain various results. For example, in Fig. \ref{FIG:12}(a), the phase mismatch in the order of $10^{-2}$ does not bring an obvious difference to the coordinated pattern demonstrating the mechanical energy steps and their corresponding pulse train. Compared to the drive-frequency match condition $|\Delta_1-\Delta_2|=\omega_m$, which demands fitting two driving fields to a quantity of another subsystem (the frequency $\omega_m$ of the mechanical oscillator), phase match for these two fields is much less rigorous. In experimental implementation of the dynamical scenario it is therefore much easier to 
have two driving field phases unmatched within a considerable range. As the relative phase $\phi$ is increased further, the resulting dynamical patterns will become rich. Two typical situations are $\phi=\pi/2$ and $\phi=\pi$, which are shown in Fig. \ref{FIG:12}(b) and Fig. \ref{FIG:12}(c), respectively. The OPP pulses, which act to the opposite moving direction of the mechanical oscillator, cannot be highly suppressed in these situations, so that the energy $\mathcal{E}_m(t)$ displays the square-shaped evolution courses.

\section{Application}

\begin{figure}
	\centering
		\includegraphics[scale=.042]{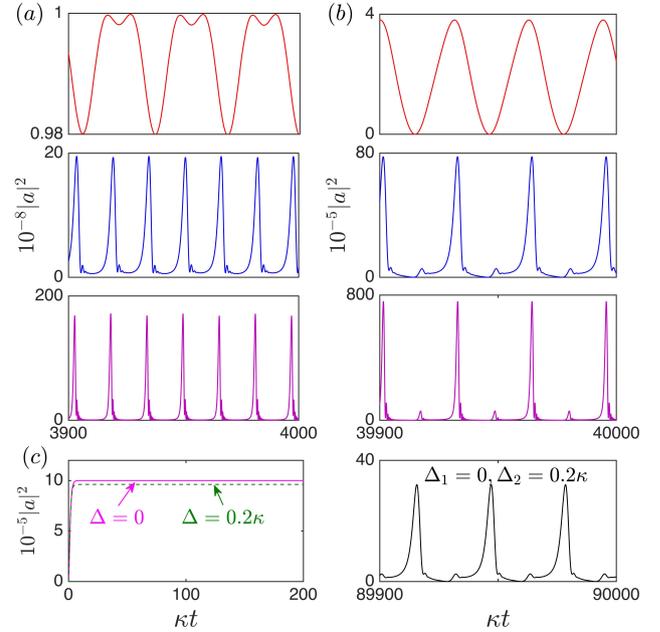}
	\caption{{\bf Cavity fields created under varied drive power.} 
(a) The samples of the asymptotically stabilized field intensity $|a(t)|^2$, driven by a single field of resonant frequency $\omega=\omega_c$. From the top to the bottom, $E/\kappa=10^4$, $5\times 10^4$, and $2\times 10^5$. 
(b) The corresponding results by two drives under the condition $|\omega_1-\omega_2|=\omega_m$, but with the drive powers being all reduced by $1000$ times ($E/\kappa=10^{2.5}$, $5\times 10^{2.5}$, and $2\times 10^{3.5}$, respectively). (c) A significantly enhanced optomechanical nonlinearity. 
Any of the single drives with the amplitude $E/\kappa=10^3$ only leads to a quasi-linear response, 
but their joint action under the condition $|\omega_1-\omega_2|=\omega_m$ creates a pulse train. 
The used OMS has $\omega_m/\kappa=0.2$, $g_m/\kappa=10^{-5}$, and $\gamma_m/\kappa=10^{-5}$.}
	\label{FIG:13}
\end{figure}

In an unresolved sideband regime $\omega_m<\kappa$, one will see a pulsed field after plugging a mechanical oscillation $X_m(t)=A\sin\omega_mt$ with a large amplitude $A$ into the first equation of Eq. (\ref{2}). Such pulsed fields were discussed with possible applications \cite{miri2018optomechanical,xu2021chip}. Especially in a recent experiment \cite{hu2021generation}, the OFC with $938$ optical comb lines was generated by pumping a nearly unresolved sideband system with a single blue-detuned drive. Based on the full dynamics of only one drive in Eq. (\ref{2}), one sees in Fig. \ref{FIG:13}(a) the transition from a field of discrete sidebands to the pulsed ones via increasing the drive power. To create the pulse train on the second row of Fig. \ref{FIG:13}(a), the power of a pump should be over $330$ mW by employing a setup with $\kappa_e=2\pi \times 25$ MHz, $\kappa_i=0.04\kappa_e$, and its resonant cavity frequency $\omega_c=2\pi \times 190$ THz, which are close to the system parameters of the experimental setup in Ref. \cite{hu2021generation}. Simply by adding another drive satisfying $|\omega_1-\omega_2|=\omega_m$, the power for generating a pulse train can be reduced by at least $1000$ times as shown in Fig. \ref{FIG:13}(b). 

In Fig. \ref{FIG:13}(c), the pump power is reduced to the level such that the system under a single drive only exhibits a quasi-linear response of approximately steady field intensity, similar to a previous experiment (that setup has $\omega_m\sim 0.001\kappa$) \cite{doolin2014nonlinear}. However, when we apply two drive tones matching their difference with the mechanical frequency to the same OMS, a pulse train will surprisingly emerge with their peaks higher than the stationary field intensity realized under either of the individual drive tones alone. The pump power to obtain the pulse train will be about $130$ $\mu$W given the above-mentioned setup close to the experimental one in Ref. \cite{hu2021generation}. Equal-distanced and approximately identical pulses are generated; for the example in Fig. \ref{FIG:5}(a), the variation of the pulse front is within the order of $\kappa \delta t\sim 0.1$ after the constantly self-adjusted pulse action over the range $\kappa \Delta t=3000$. Moreover, the frequency span of the generated OFC can be widened quickly by increasing the drive power.  

\section{Experimental feasibility and possible extensions}

Theoretically, any OMS, which has a sufficiently high mechanical quality factor $Q=\omega_m/\gamma_m$ together with a mechanical frequency $\omega_m$ lower than the cavity field damping rate $\kappa$ by a certain amount, can realize the described phenomena. The most essential requirement is to well keep the difference between two drive tones. In experiment 
the resonant cavity frequency will drift from the original $\omega_c$ due to changed temperature in operation \cite{hu2021generation}, thus adding a common detuning $\delta$ to the driving fields. However, the same phenomenon still exists after a parallel shift in 
Eq. (\ref{shift}),  
With the available techniques \cite{verlot2010backaction,shen2021dissipatively}, the frequency difference between two driving lasers 
can be kept to a good extent. Two other factors, the thermal perturbation at room temperature and the phase mismatch between two driving fields, are not important to the dynamical process as we have clarified in Sec. 5.2 and Sec. 5.3. Because of an enhanced nonlinearity, the required ratio $E/\kappa=\sqrt{2\kappa_eP/[(\kappa_e+\kappa_i)^2\hbar\omega]}$ for implementing the concerned scenarios under a fixed power $P$ can be much lowered, thus dispensing with the necessity for very high cavity finesse ($\kappa_i$ can be higher).       

Upon entering the resonance, the system operates in a stable pattern but in a dynamical instability with the mechanical amplitude approximately proportional to $\sqrt{\kappa nT}$ after $n$ mechanical oscillation periods. In the extreme situation that the mechanical oscillation amplitude becomes compatible with $\lambda_c/2=\pi c/\omega_c$, more cavity field modes neighboring the original one $a=a_0$ with its frequency $\omega_c=cN\pi/L$ ($N$ is the node number of the standing wave in the cavity) will be excited \cite{gao2015self}, and the potential in 
Eq. (\ref{potential}) will be modified to 
\begin{eqnarray}
&&V_{int}=\hbar\left\{\frac{c(N+n)\pi}{L+x_m}-\frac{c(N+n)\pi}{L}\right\}|a_n|^2\nonumber\\
&&=\hbar\left\{\frac{\omega_c+\frac{cn\pi}{L}}{1+\eta_mX_m}-(\omega_c+\frac{cn\pi}{L})\right\}|a_n|^2
\label{high-amplitude}
\end{eqnarray}
for each excited cavity field mode $a_n$ of the system in Fig. \ref{FIG:1}(a), where $n=0,\pm 1,\pm 2,\cdots$, and $\eta_m=\sqrt{\frac{\hbar}{m\omega_m}}/L$. This multi-mode extension for a previously studied scenario under a single drive \cite{poot2012backaction} is elaborated in Ref. \cite{gao2015self}, which shows that its important features (including the pulsed cavity field corresponding to the saw-tooth-edged ellipses in the limit cycles of the mechanical motion) are well preserved after such extension to multiple cavity field modes. It is expected that the main features of our concerned dynamical behaviors can be also seen after a similar generalization. To be comparable with the majority of other works adopting the parameter $g_m$ and avoid introducing a new parameter $\eta_m$ from Eq. (\ref{high-amplitude}), we work under the condition $|x_m|\ll L$ for Eq. (\ref{2}), which can be valid for a sufficiently long duration of time to many realistic experimental setups different from the one in Fig. \ref{FIG:1}. For example, the ratio $g_mE/\kappa^2$ in Fig. \ref{FIG:1}(d) can be lowered to $0.1$ when realizing the dynamical scenario with $\omega_m=0.1\kappa$. Given the parameters $g_m/\kappa=10^{-5}$ and $E/\kappa=10^{4}$ for such a setup, one still has a mechanical oscillation amplitude less than $1$ nm after $\kappa t=10^5$, considering the zero-point fluctuation amplitude $\sqrt{\hbar/(m\omega_m)}\sim 0.1$ fm and the size $L\sim 30$ $\mu$m of an available micro-cavity \cite{hu2021generation}.  

Another possible extension to multi-mode OMSs is through adding more mechanical elements or more optical cavities, similar to some previously studied systems in, for examples. Refs. \cite{bhattacharya2008multiple, metzger2008self,lin2010coherent,nielsen2017multimode, chen2017entanglement, wang2019breaking, lai2020nonreciprocal,kohler2020simultaneous,lai2020tunable,xie2020pt,lai2021significant,chen2022cooling,lai2022noise,lai2022tripartite}. This type of extensions may bring about the synchronization between the mechanical elements \cite{colombano2019synchronization, sheng2020self, wang2021passive} in addition to a field-oscillator synchronization, as well as other nontrivial phenomena such as the sudden transitions between dynamical patterns \cite{lin2021catastrophic}. Further researches will be undertaken to unveil the richer nonlinear dynamical behaviors that may exist in multi-mode systems. 

\section{Conclusion}
We have presented a detailed study on a special dynamical process in unresolved-sideband OMSs and its related issues.
Both aspects of the illustrated dynamical scenario are highly meaningful to the current researches. First, the mechanical energy exhibiting repeatedly uniform jumps provides an example that even two totally different types of motion can be synchronized by tuning a control parameter. Such a unique energy transfer brings something new to the research fields of nonlinear resonance \cite{rajasekar2016nonlinear} and synchronization \cite{pikovsky2002synchronization,boccaletti2018synchronization}. On the other side, there exists a mechanism to enhance the nonlinearity under a condition of drive-frequency match, allowing possible applications at low levels of pump power. A prospect of the wider use of the OMSs of unresolved sideband, which are less demanding in their fabrications, may be real in view of their dynamics depicted here.

\section*{Acknowledgements}

We thank Dr. Zhen Shen, Dr. Yan-Lei Zhang, Dr. Ming Li, and Dr. Luis Mart\'{i}nez for the helpful discussions on the relevant experimental issues. This work was supported by National Natural Science Foundation of China (11574093), Natural Science Foundation of Fujian Province (2020J01061), ANID Fondecyt Regular (1221250), and Fondo de Iniciaci\'{o}n de Universidad Mayor (PEP I-2019021).



\bibliographystyle{vancouver}

\bibliography{escalera-regular}

\end{document}